\documentclass[aps,prl,reprint]{revtex4-1}  

\usepackage[T1]{fontenc}
\usepackage{graphicx}
\usepackage{amssymb}

\begin{document}

\title{Atomic-scale sensor for charge- and atomic-lattice-dynamics on surfaces}

\author{Giulia Serrano} 

\author{Stefano Tebi} 

\author{Stefan Wiespointner-Baumgarthuber} 

\author{Stefan M\"{u}llegger}
\email[Corresponding author: ]{stefan.muellegger@jku.at}

\author{Reinhold Koch} 
\affiliation{Institute of Semiconductor and Solid State Physics, Johannes Kepler University Linz, 4040 Linz, Austria.} 

\newcommand{\didv}{$\mathrm{d}I/\mathrm{d}V$\,}

\begin{abstract}
We present here a powerful method providing simultaneous atomic spatial and nanosecond temporal resolution for investigating dynamics and structure on the atomic scale, in general. 
We reveal the dynamic reorganization of surface (ad)atoms induced by radio frequency alternating charging and decharging of a metal. 
Our method utilizes taylor-made nano-fabricated two-dimensional islands of physisorbed argon atoms, acting as motion sensors, probed by a radio frequency low-temperature scanning tunneling microscope. 
\end{abstract}

\maketitle

Charge oscillations forced by voltages and currents alternating at radio frequency (rf) play a dominant role in practically all electronic devices used in our daily life. 
Such charge oscillations are known to excite plasmons in the quasi-two-dimensional skin layer at the surface of metals \cite{Nagao2001,Silkin2005,Diaconescu2007}. 
In addition, due to electrostriction \cite{Smith2015}, a periodically modulated electric ($E$) field gives rise to forced small-amplitude mechanical oscillations of the atomic lattice, $E$-fields of $10^6$\,V/m typically generating strain of 0.1~\% \cite{Weissmuller2003}. 
Such mechanical oscillations are utilized as surface acoustic waves in piezoelectric biosensors \cite{Lange2008} and frequency filters for telecommunication \cite{Morgan2007}; they enable the tuning of electronic and magnetic properties of materials \cite{Gleiter2001} as well as the operation of artificial muscles for robots and artificial limbs \cite{Weissmuller2003};  
ultrasonic irradiation facilitates to control spin--lattice relaxation times and peak widths in nuclear magnetic resonance \cite{Homer2007}. 
For measuring strain, macroscopic techniques such as x-ray diffraction \cite{Weissmuller2003} or cantilever beam techniques \cite{Koch2010} are well established. 
Detection of picometer-scale mechanical amplitudes of the surface atomic lattice has been achieved up to 1\,GHz by an rf-modified scanning tunneling microscope \cite{Voigt2002}. 
Near-field microwave microscopy has demonstrated the electrodynamical response of the material on length scales far shorter than the free-space wavelength of the microwave \cite{Anlange2007}. 
However, a direct real-space detection and imaging of the surface atomic lattice and its dynamics has remained elusive, to date, due to the lack of simultaneous spatial and temporal resolution of the detection method applied. 
We have developed a method that circumvents these difficulties. 

\begin{figure}
\includegraphics[width=8.5cm]{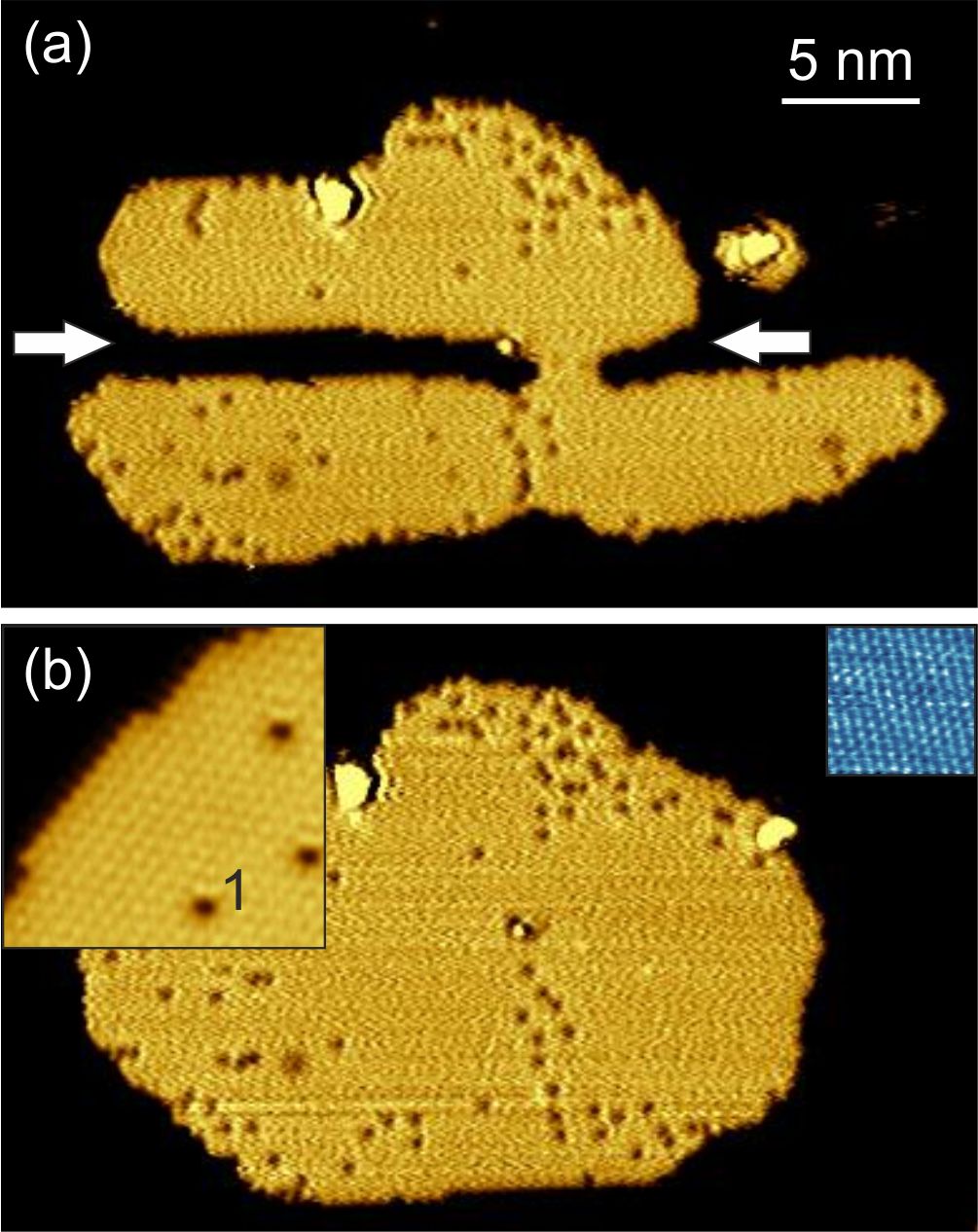}
\caption{\label{fig:sensor} (a) Nanometer-sized motion sensor: nano-fabricated 2D-island of Ar on Ag(111) imaged by STM at 5\,K ($53\times30$\,nm$^2$, $z$-scale: 200\,pm, $+0.4$\,V, 70\,pA); arrows mark lateral channels fabricated by cutting-out from the island by dc-STM manipulation. (b) Same 2D-island as in (a) before the nano-fabrication step; left inset: magnified view ($6.5\times6.5$\,nm$^2$) 
revealing atomic resolution on the Ar 2D-island as well as single Ar vacancies (labeled~1); right inset: atomic-resolution image of the Ag(111) substrate ($3\times3$\,nm$^2$, 1.2 nA, $-30$\,mV).}
\end{figure}

Here we investigate with nanometer spatial- and nanosecond temporal resolution the impact of radio-frequency alternating electric charging and de-charging of a metal on its atomic surface structure. 
In particular, we reveal the dynamic reorganization of surface (ad)atoms with a time constant of 147~ns induced by applying a 2-ns-periodic rf-voltage (530~MHz) to the metal surface. 
As origin, we identify the charge-density oscillations in the metal surface skin layer. 
For detecting such dynamic processes we utilize nanometer-sized motion sensors (Fig.~\ref{fig:sensor}) consisting of nano-structured two-dimensional (2D) islands of physisorbed noble-gas atoms \cite{Konig2008}. 
Our motion sensors are shown herein to structurally transform on oscillating substrates like an rf-biased metal surface. 
Their structural transformations are shown to be powerful analytical probes for characterizing the underlying physical processes of the excitation at the nano-scale. 

Figure~\ref{fig:sensor}a shows exemplarily a typical motion sensor imaged by STM at 5\,K. 
It is based on a 2D-island of Ar on Ag(111) that exhibits a strongly non-equilibrium shape obtained after cutting lateral channels (marked by arrows) out of the 2D-island. 
This is achieved by means of dc-STM manipulation \cite{Mullegger2015b} for the controlled removal of Ar atoms at the channels (see supplementary figure\,S1). 
We demonstrate below that these channels are suitable probes for detecting atomic-scale motional dynamics of surface (ad)atoms. 
For comparison, Fig.~\ref{fig:sensor}b displays the 2D-island of (a) before the nano-structuring step, i.e. exhibiting its natural compact equilibrium shape. 
The left inset of Fig.\,\ref{fig:sensor}b displays a magnified view revealing the regular hexagonal Ar atomic lattice of the 2D-islands with Ar-Ar distance of 0.39~nm, in agreement with previous studies  \cite{Leatherman1997,Unguris1981,Konig2008}. 
A single Ar vacancy is labeled~1. 
For comparison, the atomically resolved Ag(111) lattice ($a_{111}=a_0/\sqrt{2}=0.289$\,nm) is shown in the right inset of Fig.~\ref{fig:sensor}b. 
Similar to a monolayer of Ar on Ag(111) \cite{Konig2008}, the equilibrium-shape islands are stable for >12\,h  during continuous imaging by dc-STM at sample bias voltage of $+0.4$ to $+1.3$\,V and tunneling current of 50--200~pA. 
More importantly here, also the motion sensors, i.e. the non-equilibrium 2D-islands (Fig.~\ref{fig:sensor}a), are found to be longterm stable against dc-imaging by STM at 5\,K (see supplementary figure\,S2). 

\begin{figure}
\includegraphics[width=8.4cm]{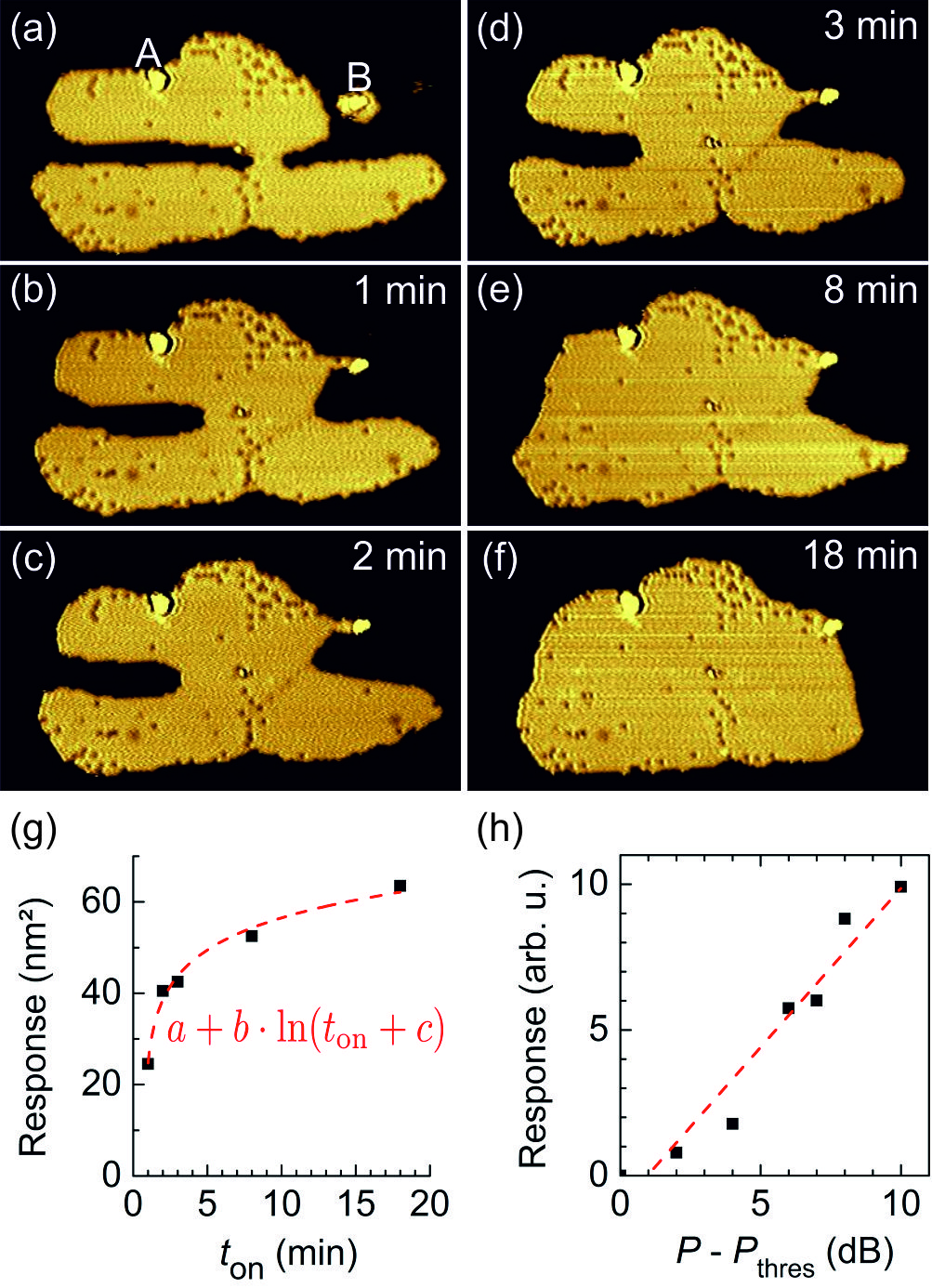}
\caption{\label{fig:response} Response of motion sensor to cw rf-excitation. (a) Sensor island before rf-excitation imaged by STM ($53\times30$~nm$^2$, $+0.4$\,V, 70\,pA). (b)--(f) Same sensor island as in (a) after successive cw rf-excitation (530\,MHz, $P_\mathrm{thres}+4$\,dB) for accumulating on-time, $t_\mathrm{on}$ (see labels). (g,h) Dependence of the magnitude of sensor response on $t_\mathrm{on}$; line: numerical fit, function given in red. (h) Dependence on rf power $P$; line: numerical linear fit.}
\end{figure}

For investigating the impact of rf alternating electric charging and de-charging of the metal on its atomic surface structure, we have connected the Ag(111) sample to the output of an rf generator in parallel to the dc sample-voltage source. 
We set a fixed generator frequency to avoid possible effects of frequency-dependent damping of the rf circuitry and to guarantee a constant rf voltage amplitude at the sample surface for all experiments presented herein.  
The frequency was $f=530$\,MHz, which means that the microwave in the Ag sample kept at 5\,K is confined to a surface skin layer \cite{Jackson1999} of thickness $\delta = \sqrt{\rho / (\mu \pi f)} < 100$\,nm (resistivity of Ag at 5\,K is $\rho=1.3\cdot10^{-11}$~$\Omega$m \cite{Chambers1952} and permeability $\mu\approx \mu_0=4\pi\cdot10^{-7}$\,Vs/Am). 
In a first step, we have applied a continuous-wave (cw) rf-voltage to the sample for varying time spans $t_\mathrm{on}$. 
Before and after each excitation, the motion sensor was imaged by dc-STM. 
Intriguingly, our motion sensors respond to the cw excitation with characteristic structural changes as evidenced in Fig.\,\ref{fig:response}. 
Excitation for $t_\mathrm{on}=1$\,min causes a gradual closing of the channels. 
This is clearly seen by comparing the images of the motion sensors before (a) and after (b) the rf excitation. 
Moreover, the large island merges with the isolated small one on the right. 
Notice the defects labeled A and B being unaffected by the rf excitation. 
The total size of the sensor (area) has remained approximately constant, indicating that no significant amount of Ar atoms is added or subtracted during $t_\mathrm{on}$. 
Obviously, channel closing proceeds via directed diffusion (displacement) of Ar atoms across the Ag surface. 
The perimeter-to-area ratio of the sensor island decreases monotonically, evidencing the non-random nature of the underlying process. 
We emphasize that the sensors do not respond to dc tunneling. 
Repeating the 1\,min cw excitation leads to a further closing of the channels (Fig.~\ref{fig:response}c), which seems to come to rest after a third excitation (Fig.~\ref{fig:response}d). 
Complete closure of the channels, however, is achieved after applying an additional 5-min (Fig.~\ref{fig:response}e) and 10-min (Fig.~\ref{fig:response}f) cw excitation. 
Finally, the motion sensor adopts a compact equilibrium-like shape similar to original Ar 2D-islands. 
By determining the total number of Ar atoms (area) displaced during the on-time of the rf-voltage, we have quantified the size of the sensor response. 
Starting from zero, with increasing $t_\mathrm{on}$  the response increases and finally saturates at very long times (Fig.~\ref{fig:response}g). 
Within the experimental range of our method, we have found a nearly linear dependence of the sensor response on the microwave power (Fig.~\ref{fig:response}h). 
It exhibits a low-power threshold of  $P_\mathrm{thres}\approx$3\,dBm generator output power, corresponding to an rf-voltage amplitude at the sample of only a few mV zero-to-peak,   considering the damping of our rf circuitry \cite{Mullegger2014b}. 

\begin{figure}
\includegraphics[width=8.4cm]{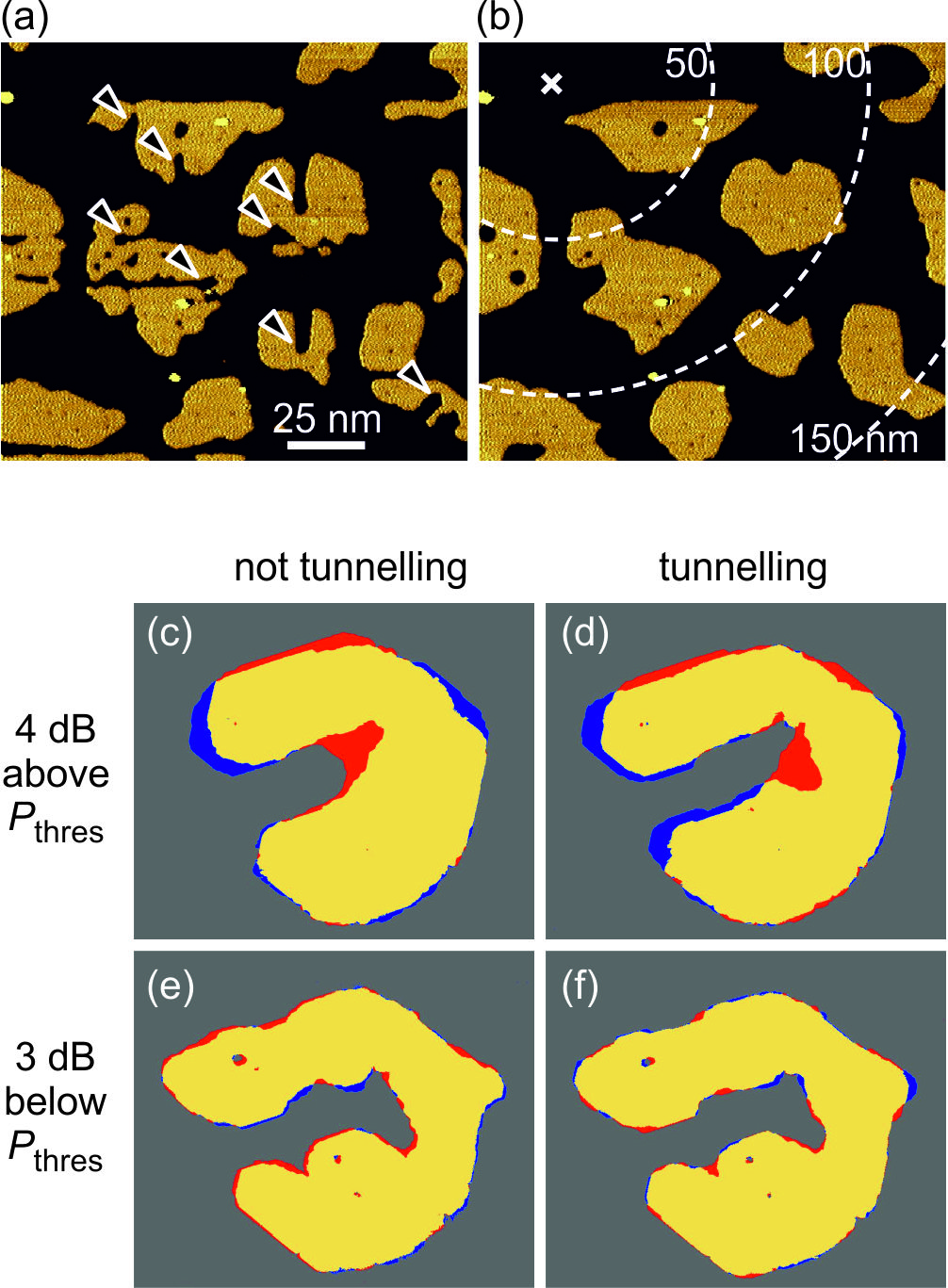}
\caption{\label{fig:tip} Sensor response is independent of STM tip position. (a,b) STM images of six sensor islands, marked by arrows, recorded before (a) and after (b) cw rf-excitation (530\,MHz, $t_\mathrm{on}=5$\,min); cross marks tunnel position of STM tip during excitation (+0.4\,V); dashed lines mark radial distance from tip position. (c-f) Difference images of sensor response obtained by subtracting STM images before and after rf-excitation ($t_\mathrm{on}=5$\,min) at different rf-power levels and tunnel conditions (see labels); red (blue) color marks positive (negative) response, i.e. accumulation (removal) of Ar atoms; unchanged sensor area is plotted in yellow.}
\end{figure}

In the experiments described so far, the STM tip was positioned over the sensor 2D-islands in tunnel contact (typically $+0.4$\,V, 100\,pA) during the rf-excitation. 
To minimize STM-tip effects, we carefully checked the tip state during the experiments and have repeated them with different tips (tip formings). 
It has turned out, however, that the position of the STM tip during rf excitation is irrelevant for the sensor response. 
Figure\,\ref{fig:tip}a shows sensor islands with eight fabricated channels in both horizontal and vertical direction, marked by arrows. 
After rf-excitation for $t_\mathrm{on} = 5$\,min all of them have responded (Fig.\,\ref{fig:tip}b), although during the excitation the STM tip was placed over the pristine substrate several tens of nanometers away from the sensors (tip position marked by cross). 
Apparently, the sensor response is based on a "non-local" mechanism, i.e. independent of the close-up range of the tip apex, and isotropic in the surface plane (see Fig.\,\ref{fig:tip}b). 
We have confirmed the sensor response up to a surface area of $400\times400$\,nm$^2$ by manual piezo control (limited by the scan range of our LT-STM instrument). 

Even more intriguing are experiments performed at  non-tunneling conditions, where the STM tip was perpendicularly retracted by $\approx$200\,nm away from the Ag(111) surface for suppressing electron tunneling;  
the local dc $E$-field between the STM tip apex and the sample is decreased by a factor of $\approx$\,200. 
For better clarity the respective results, Figs.\,\ref{fig:tip}c--f, are displayed as difference images, where red (blue) color marks sensor area where Ar atoms have been accumulated (removed) by the rf-excitation. 
The sensor response at non-tunneling  conditions, Fig.\,\ref{fig:tip}c, is clearly revealed by the channel closing after $t_\mathrm{on}=5$\,min (rf power level was 4\,dB above $P_\mathrm{thres}$). 
It is almost indistinguishable from the response at tunneling conditions (Fig.\,\ref{fig:tip}d). 
This finding clearly evidences that sensor response is independent of tunneling electrons as well as the magnitude of the dc $E$-field between tip and sample. 
Notice that edge diffusion is observed at the outermost Ar atomic row of the sensor 2D-islands in all our experiments independent of rf excitation (for details see supplementary information). 
Repeating the experiments at a decreased power level of 3\,dB below $P_\mathrm{thres}$ results in zero response (no channel closing) at both non-tunneling (Fig.\,\ref{fig:tip}e) and tunneling conditions (Fig.\,\ref{fig:tip}f). 
The respective decrease of power corresponds to a 50\%-decrease of the rf $E$-field amplitude. 
This result is indeed surprising: 
There is no response in (f), although it has at least 100 times larger $E$-field compared to (c). 
Obviously, sensor response depends on the rf-power level, but is uncorrelated with the strength of the rf $E$-field between STM tip and junction.
Our findings therefore contradict a 'simple' $E$-field effect. 
Notice that our argumentation holds independent of the precise value of the rf-voltage amplitude at the tunneling junction, which is not precisely known.

The results of the non-tunneling experiments clearly rule out (local) Joule heating at the tunnel junction, which relies on the flow of electric current \cite{Jackson1999}, as origin of the sensor response. 
Heating of the sample by microwave radiation is ruled out, because the sensor response happens in the (reactive) near-field of the sample surface, where emission of radiant energy is known to be negligible. 
Heating of the sample by absorption of electromagnetic energy in the cabling and sample crystal is ruled out because we observe negligible warming of the sample (e.g., only $<0.2$\,K upon 5\,min of cw rf-excitation).

The presented experiments suggest that 2D-islands of Ar on Ag(111) act as motion sensors responsive to the dynamic reorganization of surface (ad)atoms. 
The reorganization is caused by additional elementary diffusion processes of the sensor's atoms induced by the rf alternating electric charging and de-charging of the metal skin layer upon applying an rf-voltage to the sample. 
The additional diffusion is absent at dc-voltage conditions at 5\,K, where only edge-diffusion is observed (i.e. diffusion along the same atomic row starting off from a kink site). 
Edge diffusion alone cannot explain the response, because channel filling requires Ar atoms to move out of an edge row to form a new edge, one row in front of the old one. 
We estimate the respective energy barrier of this additional diffusion step to be on the order of $0.1$\,meV, based on our observation that 60\,min of heating the sensors to 10\,K ($k_\mathrm{B}T=0.86$\,meV) causes a similarly large sensor response as 5\,min of 530~MHz rf-excitation with power of 4\,dB above $P_\mathrm{thres}$ at 5\,K ($k_\mathrm{B}T=0.43$\,meV). 
Still, fundamental questions remain: What causes the additional diffusion process? What is the role of the Ag substrate atoms? 
In the following we discuss the underlying physical mechanism. 

\begin{figure}
\includegraphics[width=8.4cm]{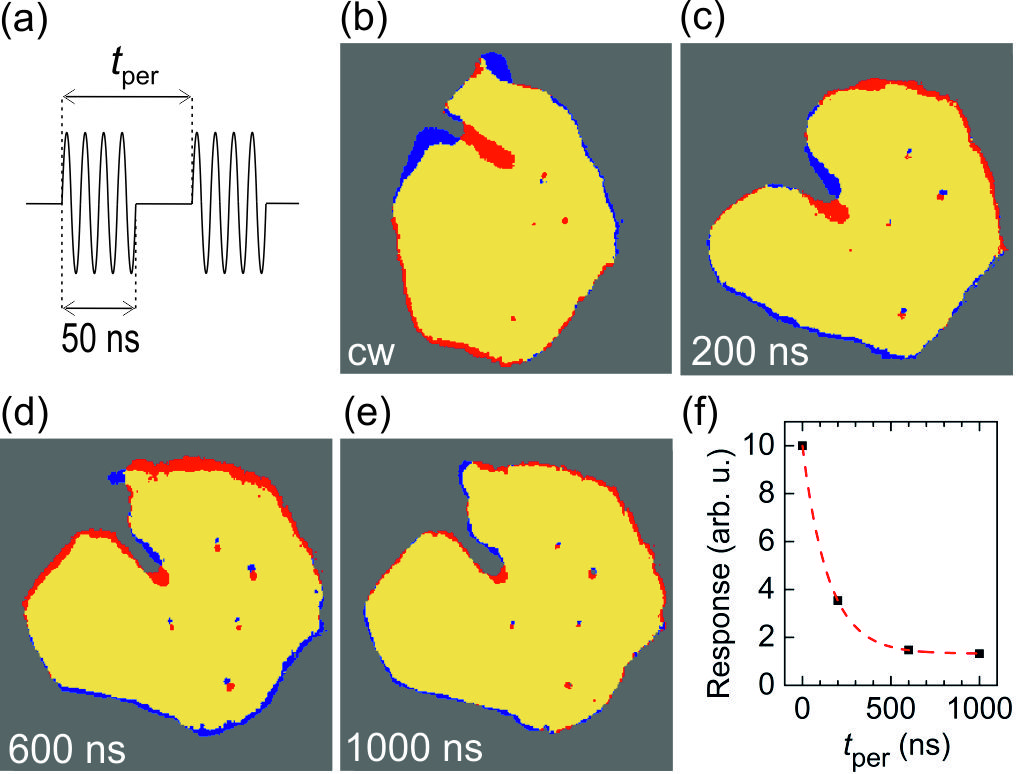}
\caption{\label{fig:puls} Sensor response to pulsed rf-excitation. (a) Schematics illustrating the periodicity $t_\mathrm{per}$ of periodic rf pulses. (b--e) Difference images of sensor response to rf excitation (530\,MHz, $P_\mathrm{thres}+7$\,dB) at non-tunneling conditions; red (blue) color marks accumulation (removal) of Ar atoms; unchanged sensor area is plotted in yellow; STM tip is positioned 500~nm away from the island center; (b) after cw-excitation for 1\,min; (c)-(e) after excitation by pulse train of $1.2\times10^9$ periodic 50~ns-pulses (equivalent of 1~min total rf on-time) with different periodicity of 200, 600, and 1000~ns. (f) Dependence of the magnitude of response on $t_\mathrm{per}$; red line: numerical fit $\propto \mathrm{exp}(t_\mathrm{per}/\tau)$.} 
\end{figure}

To gain further insight, we have investigated the sensor response to pulsed rf-excitation. 
We have obtained practically the same results for pulsed experiments at tunneling and non-tunneling conditions; for brevity we show and discuss herein only the results at non-tunneling conditions.   
We have applied pulse trains consisting of periodic 50\,ns-pulses of frequency 530\,MHz with different values of pulse period varying between $t_\mathrm{per}=200$ and 1000\,ns  (Fig.\,\ref{fig:puls}a). 
Each pulse train contained the same total number of $1.2\times10^9$ pulses, equivalent of a total on-time of 1\,min of the rf-excitation. 
Compared to cw-excitation for 1\,min (Fig.\,\ref{fig:puls}b), pulsed excitation yields a significantly smaller response (Fig.\,\ref{fig:puls}c); increasing the pulse period decreases the response further (Figs.\,\ref{fig:puls}d,e). 
This finding indicates that the dynamic processes underlying the sensor response exhibit a time constant $\tau$ with a value similar to the period of pulses applied. 
A quantitative evaluation of the $t_\mathrm{per}$-dependence of the response is displayed in Fig.\,\ref{fig:puls}f. 
Numerical fitting an exponential function yields a value of $\tau=147$\,ns for the time constant of the sensor response. 
This value is about six orders of magnitude larger than typical decay times of excited (surface) plasmons \cite{Moresco1997}, surface-state electrons \cite{Kliewer2000}, and (surface) phonons \cite{Groeneveld1995} on Ag(111). 
It seems more likely that $\tau$ belongs to the (collective) mechanical excitations of Ar atoms involved in the restructuring of the sensors. 
Similar decay times were observed for the collective vibrations of ensembles of small physisorbed molecules on Au(111) \cite{Mullegger2014a}. 

Our experimental results obviously rule out several processes as origin of the sensor response: tunnel current, the electric field between STM tip and sample, local Joule heating, radiative heating and heating by absorption of electromagnetic energy in the cabling and/or sample crystal.  
These findings indicate that sensor response is caused by processes in the Ag sample induced by the charging and decharging at radio frequency. 
In other words, we identify the periodic-in-time deviation from charge neutrality in the skin layer of the sample as origin of the observed sensor response. 
 
Since the dynamics of electrons and atomic lattice are known to be closely related to each other on time scales of $>1$\,ns, studied herein, a manifold of different effects is expected to occur simultaneously, contributing to sensor response:  
(i) Electrostriction mechanically strains the surface atomic lattice \cite{Weissmuller2003,Mullegger2015c} affecting the bonding geometry; reversible mechanical strain as large as 0.15\% has been observed in charged nanoporous Pt samples \cite{Weissmuller2003}, corresponding to mechanical stress of about 1\,GPa. 
Hence, the application of an rf voltage is expected to enforce mechanical vibrations of the surface atomic lattice at the same frequency (here 530\,MHz) with amplitude $\Delta z$, contributing to the measured dc tunnel current via the well-known relation $I_\mathrm{tunnel}\propto e^{-z}$. 
Notice, that the rf-induced change of sample surface height  $\Delta z$ may be misinterpreted as "rf-induced contribution to dc bias voltage", "apparent shift of work function", or "apparent smearing of $E_\mathrm{Fermi}$" (at const-$I$ and const-$z$ conditions).  
(ii) Excitation of acoustic surface plasmons (ASPs),  known to exist on Ag(111) \cite{Silkin2005}, is expected to influence adsorbate dynamics (here: sensor response), since the decay of ASPs can generate (surface) phonons \cite{Pitarke2005,Diaconescu2007}. 
Our sensors promise to facilitate future experimental studies on the rf-excitation of ASPs. 
(iii) Electric polarization of the surface atoms induces repulsive electrostatic forces between neighboring atoms or even affects the van der Waals-London dispersion forces of Ar-Ar as well as Ar-Ag atoms \cite{Apostol2012}, facilitating enhanced Ar diffusion. 

In conclusion, we have demonstrated that nano-fabricated monolayer 2D-islands of Ar on Ag(111) at 5\,K act as experimental probes for detecting surface (ad)atom mechanical motion and dynamic processes. 
To showcase the strength of our method, we reveal a  dynamic reorganization of surface (ad)atoms at the Ar/Ag(111) interface, caused by radio-frequency charge-density oscillations and related electric fields induced by an external rf voltage. 
Our experimental method is expected to enable quantitative characterization of atomic structural dynamics with unprecedented detail, in general. 
This is relevant, in particular, for the study of weakly bound systems with nanometer spatial- and nanosecond temporal resolution, including monolayer solids, surfaces, (hetero)interfaces and 2D nanostructures.

\section*{Methods} 
The experiments were performed in ultrahigh vacuum (base pressure: $<10^{-10}$\,mbar) with a radio-frequency low-temperature scanning tunneling microscope \cite{Mullegger2014d} operated at 5\,K. 
It utilizes a sharp tungsten tip (electrochemically etched and thermally deoxidized above 1070\,K) as, both, imaging probe as well as movable ground-electrode against the flat sample. 
The latter is a Ag(111) single-crystal prepared by repeated cycles of Ar$^+$ ion sputtering (600\,eV) and thermal annealing at 720\,K. 
The Ag(111) sample is biased from independent rf- and dc-voltage sources for applying ac and dc $E$-fields of $10^7$--$10^9$\,V/m at its surface. 
The rf-circuit and electronics are described elsewhere \cite{Mullegger2014d,Mullegger2014b}.
After cooling the sample to 5\,K, the STM chamber was flooded for 1~min with Ar gas at a pressure of $5 \cdot 10^{-7}$~mbar yielding an Ar coverage of $\approx0.3$\,monolayers on Ag(111). 
After preparation by this procedure the Ag(111) surface is covered by 2D-islands of Ar with compact shapes and typical sizes ranging from 30 to 100\,nm. 

\section*{Acknowledgments}
\begin{acknowledgments}
We kindly acknowledge financial support of the project i958 by the Austrian Science Fund (FWF). 
\end{acknowledgments}

\section*{Author contributions}
S.\,M., R.\,K. and G.\,S. designed the experiments. 
G.\,S., S.\,T., and S.\,W.-B. conducted the experiments. 
G.\,S., S.\,T., S.\,W.-B. and S.\,M. analyzed the data. 
S.\,M., R.\,K. and G.\,S. wrote the manuscript. 
S.\,M. and R.\,K. planned and supervised the project. 
All authors discussed the manuscript. 

\section*{Author information}
The authors declare no competing financial interests. 
Correspondence and requests for materials should be addressed to stefan.muellegger@jku.at. 


%

\end{document}